# Almost unbiased estimator using Known Value of Population Parameter (s) in Sample Surveys

Rajesh Singh

S.B. Gupta

Sachin Malik

# Almost unbiased estimator using Known Value of Population Parameter (s) in Sample Surveys

Rajesh Singh[1], S.B. Gupta[2] and *Sachin Malik[3]

[1] Department of Statistics, Banaras Hindu University, Varanasi-221005, India

[2,3] Department of Community Medicine, SRMS Institute of Medical Sciences

Bareilly- 243202, India

*Corresponding Author

(rsinghstat@gmail.com, dr_sbgupta@rediffmail.com, sachinkurava999@gmail.com)

## Abstract

In this paper we have proposed an almost unbiased estimator using known value of some population parameter(s). A class of estimators is defined which includes Singh and Solanki [1] and Sahai and Ray [2], Sisodia and Dwivedi [3], Singh et. al. [4], Upadhyaya and Singh [5], Singh and Tailor [6] estimators. Under simple random sampling without replacement (SRSWOR) scheme the expressions for bias and mean square error (MSE) are derived. Numerical illustrations are given in support of the present study.

## 1. Introduction

It is well known that the precision of the estimates of the population mean or total of the study variable y can be considering improved by the use of known information on an auxiliary variable x which is highly correlated with the study variable y. Consider a finite population $U=U_1,U_2,\ldots U_N$ of N units. Let y and x stand for the variable under study and auxiliary variable respectively. Let $(y_i, x_i)$, i=1,2,..,n denote the values of the units included in a sample $s_n$ of size n drawn by simple random sampling without replacement (SRSWOR). The auxiliary information has been used in improving the precision of the estimate of a parameter (See Sukhatme et. al. [7] and the references cited there in). Out of many methods, ratio and product methods of estimation are good illustrations in this context.

In order to have a survey estimate of the population mean $\overline{Y}$ of the study character y, assuming the knowledge of the population mean $\overline{X}$ of the auxiliary character x, the well-known ratio estimator is

$$t_R = \bar{y}\frac{\overline{X}}{\bar{x}} \qquad (1.1)$$

Bahl and Tuteja [8] suggested an exponential ratio type estimator as –

$$t_{exp} = \bar{y}\exp\left[\frac{\overline{X}-\bar{x}}{\overline{X}+\bar{x}}\right] \qquad (1.2)$$

Several authors have used prior value of certain population parameter(s) to find more precise estimates. Sisodiya and Dwivedi [3], Sen [9] and Upadhyaya and Singh [5] used the known coefficient of variation (CV) of the auxiliary character for estimating population mean of a study character in ratio method of estimation. The use of prior value of coefficient of kurtosis in estimating the population variance of study character y was first made by Singh et.al.[10]. Later used by Singh and Kakaran [11] in the estimation of population mean of study character. Singh and Tailor [6] proposed a modified ratio estimator by using the known value of correlation coefficient. Kadilar and Cingi [12] and Singh et. al. [13] have suggested modified ratio estimators by using different pairs of known value of population parameter(s).

In this paper under SRSWOR, we have proposed almost unbiased estimator for estimating $\overline{Y}$.

To obtain the bias and MSE , we write

$$\bar{y} = \overline{Y}(1+e_0), \quad \bar{x} = \overline{X}(1+e_1),$$

such that

$E(e_0) = E(e_1) = 0$.

$$E(e_0^2) = f_1 C_y^2, \qquad E(e_1^2) = f_1 C_x^2, \qquad E(e_0 e_1) = f_1 \rho C_y C_x$$

where $f_1 = (\frac{1}{n} - \frac{1}{N})$, $S_y^2 = \frac{1}{(N-1)}\sum_{i=1}^{N}(y_i - \overline{Y})^2$, $S_x^2 = \frac{1}{(N-1)}\sum_{i=1}^{N}(x_i - \overline{X})^2$,

$C_y = \frac{S_y}{Y}$, $C_x = \frac{S_x}{X}$, $K_x = \rho_{yx}\left(\frac{C_y}{C_x}\right)$, $\rho_{yx} = \frac{S_{yx}}{(S_y S_x)}$, $S_{yx} = \frac{1}{(N-1)}\sum_{i=1}^{N}(y_i - \overline{Y})(x_i - \overline{X})$

## 2. The proposed estimator

We proposed the following estimator

$$t_1 = \overline{y}\left(\frac{K_1\overline{X} + K_2K_3}{K_1\overline{x} + K_2K_3}\right)^{\alpha} \tag{2.1}$$

The Bias and MSE expressions of the estimator $t_1$ up to the first order of approximation are, respectively, given by

$$B(t_1) = \overline{Y}f_1C_x^2\left[\frac{\alpha(\alpha+1)V_1^2}{2} - \alpha V_1K_x\right] \tag{2.2}$$

$$MSE(t_1) = \overline{Y}^2 f_1\left[C_y^2 + C_x^2\left(\alpha^2V_1^2 - 2V_1\alpha K_x\right)\right] \tag{2.3}$$

Following Singh and Solanki [1], we propose the following estimator

$$t_2 = \overline{y}\left\{2 - \left(\frac{\overline{x}}{\overline{X}}\right)^{\beta}\exp\left[\lambda\left(\frac{(K_4\overline{X} + K_5) - (K_4\overline{x} + K_5)}{(K_4\overline{X} + K_5) + (K_4\overline{x} + K_5)}\right)\right]\right\} \tag{2.4}$$

The Bias and MSE expression's of the estimator $t_2$ up to the first order of approximation are, respectively, given by

$$B(t_2) = \overline{Y}f_1C_x^2\left[\frac{\lambda V_2\beta}{2} - \frac{\beta(\beta-1)}{2} - \frac{\lambda(\lambda+2)V_2^2}{8} - \beta K_x + \frac{\lambda V_2K_x}{2}\right] \tag{2.5}$$

$$MSE(t_2) = \overline{Y}^2 f_1\left[C_y^2 + C_x^2\left(\beta^2 + \frac{\lambda^2V_2^2}{4} - \beta\lambda V_2\right) - 2K_xC_x^2\left(\beta - \frac{\lambda V_2}{2}\right)\right] \tag{2.6}$$

$\alpha$, $\lambda$ and $\beta$ are suitable chossen constants. Also $K_1, K_3, K_4, K_5$ are either real numbers or function of known parameters of the auxiliary variable x such as $C_x$, $\beta_2(x)$, $\rho_{yx}$ and $K_x$. $K_2$ is an integer which takes values +1 and -1 for designing the estimators and

$$\left.\begin{aligned} V_1 &= \frac{K_1\bar{X}}{K_1\bar{X}+K_2K_3} \\ V_2 &= \frac{K_4\bar{X}}{K_4\bar{X}+K_5} \end{aligned}\right\}$$

We see that the estimators $t_1$ and $t_2$ are biased estimators. In some applications bias is disadvantageous. Following these estimators we have proposed almost unbiased estimator of $\bar{Y}$.

**3. Almost unbiased estimator**

Suppose $t_0 = \bar{y}$ $t_1 = \bar{y}\left(\frac{K_1\bar{X}+K_2K_3}{K_1\bar{x}+K_2K_3}\right)^{\alpha}$ $t_2 = \bar{y}\left\{2-\left(\frac{\bar{x}}{\bar{X}}\right)^{\beta}\exp\left[\lambda\left(\frac{(K_4\bar{X}+K_5)-(K_4\bar{x}+K_5)}{(K_4\bar{X}+K_5)+(K_4\bar{x}+K_5)}\right)\right]\right\}$

Such that $t_0$, $t_1$, $t_2 \in W$, where W denotes the set of all possible estimators for estimating the population mean $\bar{Y}$. By definition, the set W is a linear variety if

$$t_p = \sum_{i=0}^{3} w_i t_i \in W \tag{3.1}$$

Such that, $\sum_{i=0}^{3} w_i = 1$ and $w_i \in R$ (3.2)

where $w_i(i = 0,1,2,3)$ denotes the constants used for reducing the bias in the class of estimators, H denotes the set of those estimators that can be constructed from $t_i(i = 0,1,2,3)$ and R denotes the set of real numbers.

Expressing (3.1) in terms of e's, we have

$$t_p = \bar{Y}\left[1+e_0+w_1\left(\frac{\alpha(\alpha+1)V_1^2e_1^2}{2}-\alpha V_1e_1-\alpha V_1e_0e_1\right)\right.$$

$$+w_2\left(-\beta e_1-\frac{\beta(\beta-1)e_1^2}{2}+\frac{\lambda V_2 e_1}{2}+\frac{\lambda V_2\beta e_1^2}{2}-\frac{\lambda(\lambda+2)V_2^2e_1^2}{8}-\beta e_0e_1+\frac{\lambda V_2 e_0 e_1}{2}\right)\Bigg] \tag{3.3}$$

Subtracting $\overline{Y}$ from both sides of equation (3.3) and then taking expectation of both sides, we get the bias of the estimator $t_p$ up to the first order of approximation, as

$$B(t_p)=\overline{Y}f_1w_1C_x^2\left(\frac{\alpha(\alpha+1)V_1^2}{2}-\alpha V_1K_x\right)+\overline{Y}f_1w_2C_x^2\left(\frac{\lambda V_2\beta}{2}-\frac{\beta(\beta-1)}{2}-\frac{\lambda(\lambda+2)V_2^2}{8}\right.$$

$$\left.-\beta K_x+\frac{\lambda V_2K_x}{2}\right) \tag{3.4}$$

From (3.3), we have

$$(t_p-\overline{Y})=\overline{Y}\left[e_0-w_1\alpha V_1e_1-w_2\left(\beta e_1+\frac{\lambda V_2e_1}{2}\right)\right] \tag{3.5}$$

Squaring both sides of (3.5) and then taking expectation, we get the MSE of the estimator $t_p$ up to the first order of approximation, as

$$MSE(t_p)=\overline{Y}^2f_1\left[C_y^2+C_x^2\left(Q^2-2QK_x\right)\right] \tag{3.6}$$

Which is minimum when

$$Q=K_x \tag{3.7}$$

Where $Q=w_1\alpha V_1+w_2\left(\beta-\frac{\lambda V_2}{2}\right)$ (3.8)

Putting the value of $Q=K_x$ in (3.6) we have optimum value of estimator as $t_p$ (optimum).

Thus the minimum MSE of $t_p$ is given by

$$\min.MSE(t_p)=\overline{Y}^2f_1C_y^2\left(1-\rho_{yx}^2\right) \tag{3.9}$$

Which is same as that of traditional linear regression estimator.

from (3.2) and (3.8), we have only two equations in three unknowns. It is not possible to find the unique values for $w_i$'s, 1=0,1,2. In order to get unique values of $w_i$'s, we shall impose the linear restriction

$$\sum_{i=0}^{2} w_i B(t_i) = 0 \tag{3.10}$$

where $B(t_i)$ denotes the bias in the $i^{th}$ estimator.

Equations (3.2), (3.8) and (3.10) can be written in the matrix form as

$$\begin{bmatrix} 1 & 1 & 1 \\ 0 & \alpha V_1 & \beta - \frac{\lambda V_2}{2} \\ 0 & B(t_1) & B(t_2) \end{bmatrix} \begin{bmatrix} w_0 \\ w_1 \\ w_2 \end{bmatrix} = \begin{bmatrix} 1 \\ k_x \\ 0 \end{bmatrix} \tag{3.11}$$

Using (3.11), we get the unique values of $w_i$'s, 1=0,1,2 as

$$\left.\begin{aligned} w_0 &= \frac{\alpha V_1[\alpha V_1 A_2 - A_1 X_1] - X_1 K_x A_1 - X_2 \alpha V_1[\alpha V_1 A_2 - A_1 X_1] - \alpha V_1 K_x A_1}{\alpha V_1[\alpha V_1 A_2 - A_1 X_1]} \\ w_1 &= \frac{X_1 K_x A_1}{\alpha V_1[\alpha V_1 A_2 - A_1 X_1]} + X_2 \\ w_2 &= \frac{K_x A_1}{[\alpha V_1 A_2 - A_1 X_1]} \end{aligned}\right\}$$

where,

$$\left.\begin{aligned} A_1 &= \frac{\alpha(\alpha+1)V_1^2}{2} - \alpha V_1 K_x \\ A_2 &= \frac{\lambda V_2 \beta}{2} - \frac{\beta(\beta-1)}{2} - \frac{\lambda(\lambda+2)V_2^2}{8} - \beta K_x + \frac{\lambda V_2 K_x}{2} \\ X_1 &= A_1\left[\beta - \frac{\lambda V_2}{2}\right] \\ X_2 &= \frac{K_x}{\alpha V_1} \end{aligned}\right\}$$

Use of these $w_i$'s, 1=0,1,2 remove the bias up to terms of order $o(n^{-1})$ at (3.1).

## 4. Empirical study

For empirical study we use the data sets earlier used by Kadilar and Cingi [12] (population 1) and Khosnevisan et. al. [14] (population 2) to verify the theoretical results.

**Data statistics:**

| **Population** | N | n | $\overline{Y}$ | $\overline{X}$ | $C_y$ | $C_x$ | $\rho_{yx}$ | $\beta_2(x)$ |
|---|---|---|---|---|---|---|---|---|
| **Population 1** | 106 | 20 | 2212.59 | 27421.7 | 5.22 | 2.10 | 0.86 | 34.57 |
| **Population 2** | 20 | 8 | 19.55 | 18.8 | 0.355 | 0.394 | -0.92 | 3.06 |

**Table 4.1** : Values of $w_i$

| $w_i$ | **Population 1** | **Population 2** |
|---|---|---|
| $w_0$ | 2.104965 | 3.692323 |
| $w_1$ | -6.48599 | 1.379436 |
| $w_2$ | 5.381022 | -4.07176 |

The percent relative efficiencies (PRE) of various estimators of $\overline{Y}$ are computed and presented in Table 4.2 below.

**Table 4.2: PRE of different estimators of $\overline{Y}$ with respect to $\overline{y}$**

| **Choice of scalars** | | | | | | | | | | | | | |
|---|---|---|---|---|---|---|---|---|---|---|---|---|---|
| $w_0$ | $w_1$ | $w_2$ | $K_1$ | $K_2$ | $K_3$ | $K_4$ | $K_5$ | $\alpha$ | $\beta$ | $\lambda$ | **Estimator** | **PRE (POPI)** | **PRE (POPII)** |
| 1 | 0 | 0 | | | | | | | | | $\overline{y}$ | 100 | 100 |
| | | | | | | | | | | | | | |
| 0 | 1 | 0 | 1 | 1 | 0 | | | 1 | | | $t_R$ | 212.80 | 24.69 |
| | | | 1 | 1 | 0 | | | -1 | | | $t_{exp}$ | 53.94 | 583.07 |
| 0 | 0 | 1 | | | | | | | 1 | 0 | $t_{1(1,0)}$ | 212.80 | 23.39 |
| | | | | | | | | | -1 | 0 | $t_{1(-1,0)}$ | 53.94 | 527.29 |
| | | | | | | 1 | 0 | | 1 | 1 | $t_{2(1,1)}$ | 143.99 | 42.93 |
| | | | | | | 1 | 0 | | 1 | -1 | $t_{2(1,-1)}$ | 306.54 | 14.63 |
| | | | | | | 1 | 0 | | 0 | 1 | $t_{2(0,1)}$ | 72.12 | 348.58 |
| | | | | | | 1 | 0 | | 0 | -1 | $t_{2(0,-1)}$ | 143.97 | 42.93 |
| $w_0$ | $w_1$ | $w_2$ | 1 | 1 | 1 | 1 | 1 | 1 | 1 | 1 | $t_P$ **Optimum** | **384.02** | **651.04** |

## 5. Proposed estimators in two phase sampling

When $\bar{X}$ is unknown, it is sometimes estimated from a preliminary large sample of size n' on which only the characteristic x is measured (for details see Singh et. al. (2007)). Then a second phase sample of size n (n < n') is drawn on which both y and x characteristics are measured. Let $\bar{x} = \frac{1}{n'}\sum_{i=1}^{n'} x_i$ denote the sample mean of x based on first phase sample of size n', $\bar{y} = \frac{1}{n}\sum_{i=1}^{n} y_i$ and $\bar{x} = \frac{1}{n}\sum_{i=1}^{n} x_i$, be the sample means of y and x respectively based on second phase of size n.

In two-phase sampling the estimator $t_p$ will take the following form

$$t_{pd} = \sum_{i=0}^{3} h_i t_{id} \in H \tag{5.1}$$

Such that, $\sum_{i=0}^{3} h_i = 1$ and $h_i \in R$ (5.2)

Where,

$$t_{0d} = \bar{y}\ ,t_{1d} = \bar{y}\left(\frac{K_1\bar{x}' + K_2K_3}{K_1\bar{x} + K_2K_3}\right)^m\ ,t_{2d} = \bar{y}\left\{2 - \left(\frac{\bar{x}}{\bar{\bar{x}}'}\right)^q \exp\left[\gamma\left(\frac{\left(K_4\bar{x}' + K_5\right) - \left(K_4\bar{x} + K_5\right)}{\left(K_4\bar{x}' + K_5\right) + \left(K_4\bar{x} + K_5\right)}\right)\right]\right\}$$

The Bias and MSE expression's of the estimator $t_{1d}$ and $t_{2d}$ up to the first order of approximation are, respectively, given by

$$B\left(t_{1d}\right) = \bar{Y}\left[\frac{m(m-1)R_1^2 f_2 C_x^2}{2} + \frac{m(m+1)R_1^2 f_1 C_x^2}{2} - m^2 R_1^2 f_2 C_x^2 + mR_1 f_3 K_x C_x^2\right] \tag{5.3}$$

$$MSE\left(t_{1d}\right) = \bar{Y}^2\left[f_1 C_y^2 + m^2 R_1^2 f_3 C_x^2 - 2mR_1 K_x f_3 C_x^2\right] \tag{5.4}$$

$$B\left(t_{2d}\right) = \bar{Y}\left[-\frac{q(q-1)f_1 C_x^2}{2} + \frac{q(q+1)f_2 C_x^2}{2} + qf_2 K_x C_x^2 + q^2 f_2 C_x^2 + f_3\gamma R_2 K_x C_x^2 + f_3\gamma R_2 q C_x^2\right] \tag{5.5}$$

$$MSE\left(t_{2d}\right) = \bar{Y}^2\left[f_1 C_y^2 + L_1^2 f_3 C_x^2\right] \tag{5.6}$$

where,

$$\left.\begin{aligned} R_1 &= \frac{K_1\bar{X}}{K_1\bar{X} + K_2K_3} \\ R_2 &= \frac{K_4\bar{X}}{2\left[K_4\bar{X} + K_5\right]} \\ L_1 &= q - \gamma A_2 \end{aligned}\right\} \tag{5.7}$$

Expressing (5.1) in terms of e's, we have

$$t_{pd} = \bar{Y}\left[1 + e_0 + w_1\left(\frac{m(m+1)R_1^2 e_1^2}{2} - mR_1 e_1 - mR_1 e_0 e_1 + mR_1 e_0 e_1 + \frac{m(m-1)R_1^2 e_1'^2}{2} + mR_1 e_0 e_1'\right)\right.$$

$$+w_2\left(-qe_1-\frac{q(q-1)e_1^2}{2}+qe'_1+q^2e_1e'_1-\frac{q(q+1)e'^2_1}{2}-\gamma R_2(e'_1-e_1)+\gamma R_2(e_0e_1-e_0e'_1)-qe_0e_1\right)\Bigg] \tag{5.8}$$

Subtracting $\overline{Y}$ from both sides of equation (5.8) and then taking expectation of both sides, we get the bias of the estimator $t_{pd}$ up to the first order of approximation, as

$$B(t_{pd})=\overline{Y}[B(t_{1d})+B(t_{2d})] \tag{5.9}$$

Also,

$$(t_{pd}-\overline{Y})=\overline{Y}[e_0+w_1[mR_1e'_1-mR_1e_1]+w_2(-qe_1+qe'_1-\gamma R_2e'_1+\gamma R_2e_1)] \tag{5.10}$$

Squaring both sides of (5.10) and then taking expectation, we get the MSE of the estimator $t_{pd}$ up to the first order of approximation, as

$$MSE(t_{pd})=\overline{Y}^2 f_1C_y^2+L_2^2f_3C_x^2-2L_2f_3K_xC_x^2 \tag{5.11}$$

Which is minimum when

$$L_2=K_x \tag{5.12}$$

Where $L_2=h_1mR_1+h_2(q-\gamma R_2)$ (5.13)

Putting the value of $L_2=K_x$ in (5.11), we have optimum value of estimator as $t_{pd}$ (optimum).

Thus the minimum MSE of $t_{pd}$ is given by

$$\min.MSE(t_{pd})=\overline{Y}^2C_y^2(f_1-f_3\rho_{yx}^2) \tag{5.14}$$

Which is same as that of traditional linear regression estimator.

from (5.2) and (5.13), we have only two equations in three unknowns. It is not possible to find the unique values for $h_i$'s, 1=0,1,2. In order to get unique values of $h_i$'s, we shall impose the linear restriction

$$\sum_{i=0}^{2}h_iB(t_i)=0 \tag{5.15}$$

where, $B(t_i)$ denotes the bias in the $i^{th}$ estimator.

Equations (5.2), (5.13) and (5.15) can be written in the matrix form as

$$\begin{bmatrix} 1 & 1 & 1 \\ 0 & mR_1 & q-\gamma R_2 \\ 0 & B(t_{1d}) & B(t_{2d}) \end{bmatrix} \begin{bmatrix} h_0 \\ h_1 \\ h_2 \end{bmatrix} = \begin{bmatrix} 1 \\ K_x \\ 0 \end{bmatrix} \tag{5.16}$$

Using (5.16), we get the unique values of $h_i$'s, i=0,1,2 as

$$\left.\begin{aligned} h_0 &= 1 - h_1 - h_2 \\ h_1 &= \frac{k_x}{mR_1} - \frac{N_1K_x(q-\gamma R_2)}{N_1q - mR_1N_2 - N_1\gamma R_2} \\ h_2 &= \frac{K_xN_1}{[N_1q - mR_1N_2 - N_1\gamma R_2]} \end{aligned}\right\}$$

Where,

$$\left.\begin{aligned} N_1 &= \frac{m(m-1)R_1^2f_2C_x^2}{2} + \frac{m(m+1)R_1^2f_1C_x^2}{2} - m^2R_1^2f_2C_x^2 + mR_1f_3K_xC_x^2 \\ N_2 &= \left[-\frac{q(q-1)f_1C_x^2}{2} + \frac{q(q+1)f_2C_x^2}{2} + qf_2K_xC_x^2 + q^2f_2C_x^2 + f_3\gamma R_2K_xC_x^2 + f_3\gamma R_2qC_x^2\right] \end{aligned}\right\}$$

Use of these $h_i$'s, i=0,1,2 remove the bias up to terms of order $o(n^{-1})$ at (5.1).

**Conclusion**

In this paper, we have proposed an unbiased estimator $t_p$. Expressions for bias and MSE's of the proposed estimators are derived up to first degree of approximation From theoretical discussion and empirical study we conclude that the proposed estimators $t_p$ under optimum conditions perform better than other estimators considered in the article. Double sampling version of the suggested class is also discussed along with its properties.

**Appendix A. Some members of the proposed family of estimators -**

| Some members (ratio-type) of the class $t_1$<br><br>**When** $w_0 = 0,\ w_1 = 1,\ w_2 = 0$<br><br>$\alpha = 1,$ | | | |
|---|---|---|---|
| $K_1$ | $K_3$ | **PRE'S**<br>$K_2 = 1$ | **PRE'S**<br>$K_2 = -1$ |
| 1 | $C_x$ | 212.80 | 212.82 |
| 1 | $\beta_2(x)$ | 212.60 | 213.02 |
| $\beta_2(x)$ | $C_x$ | 212.81 | 212.81 |
| $C_x$ | $\beta_2(x)$ | 212.71 | 212.91 |
| 1 | $\rho_{yx}$ | 212.81 | 212.82 |
| $N\overline{X}$ | $S_x$ | 212.81 | 212.81 |
| $N\overline{X}$ | f | 212.80 | 212.82 |
| $\beta_2(x)$ | $K_x$ | 212.60 | 213.02 |
| N | $K_x$ | 212.81 | 212.81 |
| N | 1 | 212.71 | 212.91 |
| N | $C_x$ | 212.81 | 212.82 |
| N | $\rho_{yx}$ | 212.81 | 212.81 |
| N | $S_x$ | 212.80 | 212.82 |
| N | f | 212.60 | 213.02 |
| N | g=1-f | 212.81 | 212.81 |
| N | $K_x$ | 212.71 | 212.91 |
| N | $\rho_{yx}$ | 212.81 | 212.82 |
| N | $S_x$ | 212.81 | 212.81 |
| N | f | 212.80 | 212.82 |
| N | g=1-f | 212.60 | 213.02 |
| N | $K_x$ | 212.81 | 212.81 |
| $\beta_2(x)$ | $\overline{X}$ | 212.71 | 212.91 |
| $N\overline{X}$ | $\overline{X}$ | 212.81 | 212.82 |
| N | $\overline{X}$ | 212.81 | 212.81 |
| n | $\overline{X}$ | 212.81 | 212.81 |

**Appendix B.**

| **Some members (product-type) of the class** $t_1$<br><br>**When** $w_0 = 0,\ w_1 = 1,\ w_2 = 0$<br><br>$\alpha = -1,$ | | | |
|---|---|---|---|
| $K_1$ | $K_3$ | **PRE'S**<br>$K_2 = 1$ | **PRE'S**<br>$K_2 = -1$ |
| 1 | $C_x$ | 550.91 | 501.92 |
| 1 | $\beta_2(x)$ | 646.03 | 314.33 |
| $\beta_2(x)$ | $C_x$ | 535.22 | 519.18 |
| $C_x$ | $\beta_2(x)$ | 582.35 | 91.18 |
| 1 | $\rho_{yx}$ | 466.00 | 579.15 |
| $N\overline{X}$ | $S_x$ | 528.52 | 526.06 |
| $N\overline{X}$ | f | 527.30 | 527.28 |
| $\beta_2(x)$ | $K_x$ | 510.01 | 543.74 |
| N | $K_x$ | 527.15 | 527.43 |
| N | 1 | 530.39 | 524.16 |
| N | $C_x$ | 528.52 | 526.03 |
| N | $\rho_{yx}$ | 524.41 | 530.15 |
| N | $S_x$ | 549.55 | 503.49 |
| N | f | 527.53 | 527.06 |
| N | g=1-f | 530.16 | 524.40 |
| N | $K_x$ | 524.70 | 529.87 |
| n | $\rho_{yx}$ | 520.05 | 534.38 |
| n | $S_x$ | 579.44 | 465.58 |
| n | f | 527.88 | 526.71 |
| n | g=1-f | 534.42 | 520.01 |
| n | $K_x$ | 520.77 | 533.69 |
| $\beta_2(x)$ | $\bar{X}$ | 622.76 | 146.09 |
| $N\bar{X}$ | $\bar{X}$ | 530.39 | 524.16 |
| N | $\bar{X}$ | 580.14 | 464.59 |
| n | $\bar{X}$ | 632.80 | 363.64 |

**Appendix C.**

| **Some members (product-type) of the class** $t_2$ <br> **When** $w_0 = 0,\ w_1 = 0,\ w_2 = 1$ | | |
|---|---|---|
| $K_4$ | $K_5$ | PRE'S |
| | | |

| $K_1$ | $K_3$ | $(\beta = -1, \lambda = -1)$ |
|---|---|---|
| 1 | $C_x$ | 358.00 |
| 1 | $\beta_2(x)$ | 423.38 |
| $\beta_2(x)$ | $C_x$ | 351.94 |
| $C_x$ | $\beta_2(x)$ | 357.48 |
| 1 | $\rho_{yx}$ | 324.10 |
| $N\bar{X}$ | $S_x$ | 349.09 |
| $N\bar{X}$ | F | 348.58 |
| $\beta_2(x)$ | $K_x$ | 341.45 |
| N | $K_x$ | 348.52 |
| N | 1 | 349.89 |
| N | $C_x$ | 349.09 |
| N | $\rho_{yx}$ | 347.37 |
| N | $S_x$ | 358.21 |
| N | F | 348.68 |
| N | g=1-f | 349.79 |
| N | $K_x$ | 347.49 |
| n | $\rho_{yx}$ | 345.56 |
| n | $S_x$ | 372.38 |
| n | F | 348.82 |
| n | g=1-f | 351.60 |
| n | $K_x$ | 345.86 |
| $\beta_2(x)$ | $\bar{X}$ | 345.86 |
| $N\bar{X}$ | $\bar{X}$ | 349.89 |
| N | $\bar{X}$ | 372.74 |
| n | $\bar{X}$ | 407.06 |

In addition to above estimators a large number of estimators can also be generated from the proposed estimators just by putting different values of constants $w_i$'s, $h_i$'s $K_1$, $K_2$, $K_3$, $K_4$, $K_5$, $\alpha$, $\beta$ and $\lambda$.